# Spin-wave propagation in ultra-thin YIG based waveguides

M. Collet[1], O. Gladii[2], M. Evelt[3], V. Bessonov[4], L. Soumah[1], P. Bortolotti[1], S.O Demokritov[3,4], Y. Henry[2], V. Cros[1], M. Bailleul[2], V.E. Demidov[3], and A. Anane[1,*]

[1] *Unité Mixte de Physique, CNRS, Thales, Univ. Paris-Sud, Université Paris-Saclay, 91767 Palaiseau, France*
[2] *Institut de Physique et Chimie des Matériaux de Strasbourg, UMR 7504 CNRS, Université de Strasbourg, 67034 Strasbourg, France*
[3] *Institute for Applied Physics and Center for Nanotechnology, University of Muenster, 48149 Muenster, Germany*
[4] *M.N. Miheev Institute of Metal Physics of Ural Branch of Russian Academy of Sciences, Yekaterinburg 620041, Russia*

**Spin-wave propagation in an assembly of microfabricated 20 nm thick, 2.5 μm wide Yttrium Iron Garnet (YIG) waveguides is studied using propagating spin-wave spectroscopy (PSWS) and phase resolved micro-focused Brillouin Light Scattering (μ-BLS) spectroscopy. We show that spin-wave propagation in 50 parallel waveguides is robust against microfabrication induced imperfections. Spin-wave propagation parameters are studied in a wide range of excitation frequencies for the Damon-Eshbach (DE) configuration. As expected from its low damping, YIG allows the propagation of spin waves over long distances (the attenuation lengths is 25 μm at $\mu_0 H = 45$ mT). Direct mapping of spin waves by μ-BLS allows us to reconstruct the spin-wave dispersion relation and to confirm the multi-mode propagation in the waveguides, glimpsed by propagating spin-wave spectroscopy.**

Magnonics holds the promise of realizing a spin-wave (SW) computational platform for analog and digital signal processing [1–3]. However, complex architectures that would be relevant for applications would only be possible if the SW can propagate on large enough distances without need for buffering. Thus, the necessity to use a low loss propagation medium for SW signals processing. Due to its low magnetic losses, Yttrium Iron Garnet (YIG) is one of the candidate materials and has been indeed used to realize proof of concept devices [4,5]. Furthermore, Large scale integration of magnonics circuits will require wafer scale microfabrication of YIG, which is now possible thanks to the advent of ultra-thin high quality YIG films [6–8]. In this letter, we study spin-wave propagation in an assembly of microfabricated 20-nm thick YIG waveguides using two complementary spectroscopic techniques, namely the propagating spin-wave spectroscopy (PSWS) [9–11] and phase resolved micro-focused Brillouin Light Scattering (μ-BLS) spectroscopy [12]. We observe SW propagation in parallel waveguides over distances as large as 70 μm. The obtained values of the propagation lengths are close to those expected from analytical models, which demonstrates the robustness of SW propagation in our thin YIG films against microfabrication induced imperfections. Moreover, using the phase resolved μ-BLS it was possible to reconstruct the multi-mode spin-wave dispersion relation.

A pulsed laser deposited, 20 nm thick YIG film with a Gilbert damping of $\alpha = 4\times10^{-4}$ and an effective magnetization of $\mu_0 M_{eff} = 0.213$ T has been used for this study. Details about the growth and the standard characterization of the structural properties of our YIG films can be found in d'Allivy Kelly *et al.* [6]. Different devices containing a series of 50 parallel waveguides with approximately a 2.5 μm width separated by 1.5 μm have been defined by laser-lithography (bottom panel of Fig. 1(b)). Due to miss-focusing during the lithography process not all the waveguides have the same width, a significant spread is obtained (top panel of Fig. 1(b)). Milling of the YIG film has been performed using Ar etching. Due to the insulating character of the substrate, the Ar beam was electrically neutralized [12]. The Au (200 nm)/Ti (20 nm) inductive antennas have been deposited directly on the YIG film and connected to a vector network analyzer (PSWS) or a microwave source (μ-BLS). The asymmetric U shape of the antennas consisting of a 1.5 μm wide signal line and a 10 μm wide ground line separated by a 2 μm gap allows for the excitation of spin waves over a wide range of wavevector. Three separation distances between antennas have been selected (D = 30, 50 and 70 μm).

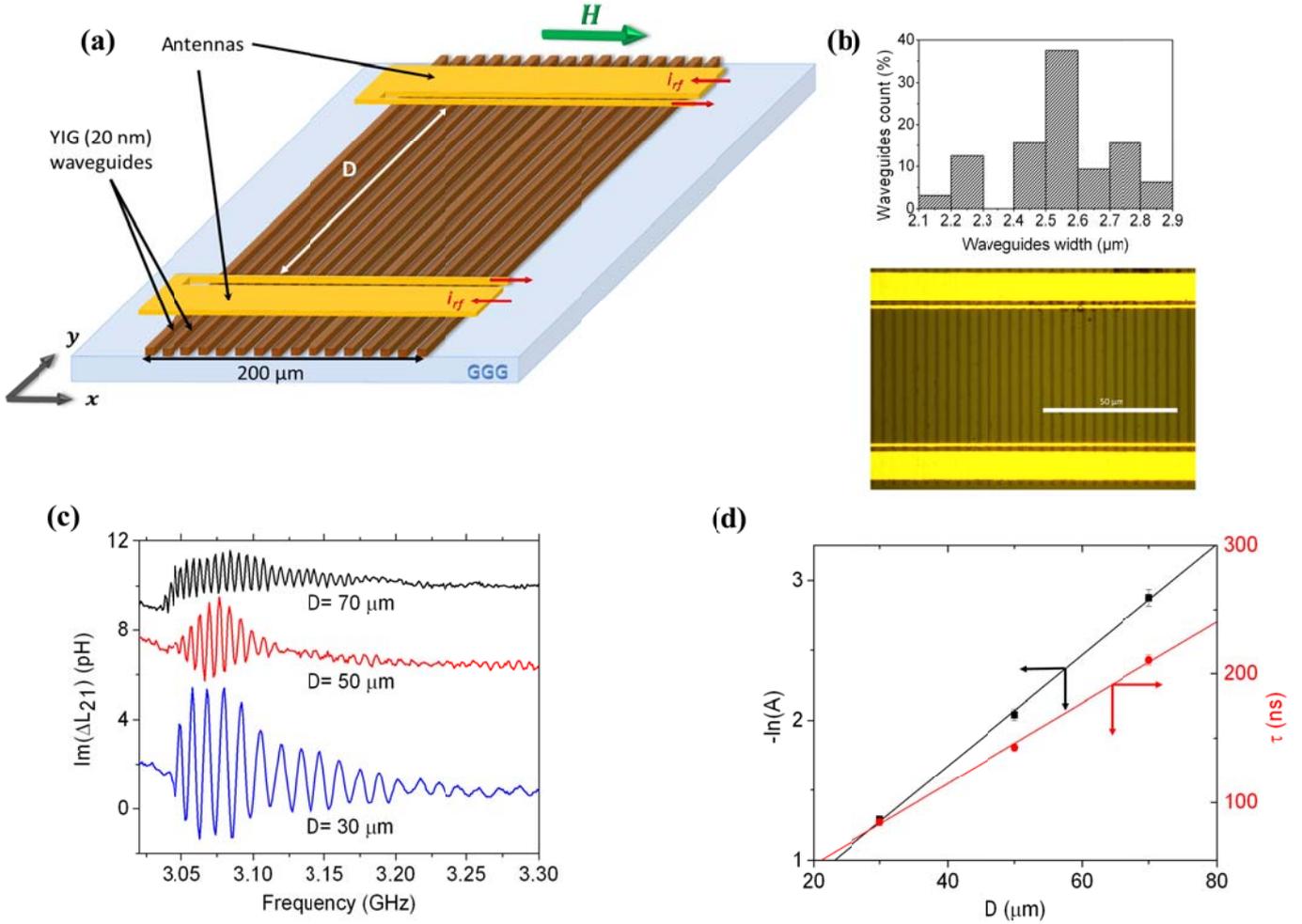

*FIG. 1. (a) Sketch of a device based on the 20 nm thick YIG film. While the total width of the device is kept constant at 200 μm, different separations between antennas: D = 30, 50 or 70 μm were used. (b) (Top panel) Histogram representing the waveguides width dispersion. (Bottom panel) Optical microscopic image of a device with a distance D = 50 μm between antennas. (c) Mutual-inductance spectra representing Im(ΔL$_{21}$) measured at $\mu_0 H$ = 45 mT for different distances between antennas D = 30, 50 and 70 μm. The data are vertically offset for clarity. (d) Dependence of the logarithm of the SW signal amplitude -ln(A) (black squares) and of the propagation time τ (red circles) on the distance D. Solid lines are linear fits of the data.*

The PSWS experiments have been performed in the Damon-Eshbach (DE) configuration, i.e., with the magnetic field applied in the film plane and perpendicular to the direction of SW propagation. The

SOLT (Short, Open, Load, Thru) calibration procedure has been used to define the microwave reference planes at the probe termination. At the input antenna, a microwave current excites SWs which then propagate in the YIG film toward the second antenna which detects the oscillating magnetic flux. From the measurements of the S-parameters, the impedance matrix $Z_{ij}$, where the indices $i$ and $j$ ($i,j$ = 1, 2) correspond to the detecting and exciting antenna, respectively, is obtained. Then, the inductance matrix $\Delta L_{ij}$ is calculated by subtracting two sets of data taken at different magnetic fields: one at the SW resonance field $H_{res}$ and the other at a reference field $H_{ref}$ far from the resonance: $\Delta L_{ij} = \frac{1}{i\omega}\left[Z_{ij}(\omega, H_{res}) - Z_{ij}(\omega, H_{ref})\right]$. From the mutual-inductance $\Delta L_{ij}$ we can extract the spin waves' propagation characteristics between the two antennas. Simultaneously, the efficiency of the coupling between the microwave circuit and the magnetic medium is monitored through the self-inductance responses $\Delta L_{ii}$ ($i$ = 1, 2).

In Fig. 1(c), we present the imaginary part of the mutual-inductance $\Delta L_{21}$ recorded at $\mu_0 H$ = 45 mT for the three distances between antennas ($D$ = 30, 50, and 70 μm). One clearly observes that both the oscillation period and the amplitude of the waveforms decrease with increasing $D$. Analyzing these two quantities, the main parameters defining the spin-wave propagation in the ensemble of YIG waveguides can be inferred. The decay of the amplitude of the transmitted signal can be expressed as $A = A_0 \exp(-(D + D_{eff})/L_{att})$, where $A = |\Delta L_{21}|^{max}/\sqrt{|\Delta L_{11}|^{max}|\Delta L_{22}|^{max}}$ is the maximum amplitude of the mutual-inductance normalized to those of the two self-inductances, $L_{att}$ is the attenuation length (corresponding to the length over which the spin-wave amplitude decreases by a factor of $e$), and $D_{eff}$ is the effective width of the antenna, which accounts for the propagation losses directly under the antenna[13]. Consequently, as shown in Fig. 1(d) (black squares), the value of the attenuation length $L_{att}$ can be deduced from the inverse of the slope, when plotting $-ln(A)$ as a function of $D$. The propagation time $\tau$, which is the inverse of the oscillation period, is related to the SW group velocity $V_g$ through $\tau = (D + D_{eff})/V_g$. Thus, by plotting $\tau$ as a function of $D$ (red circles in Fig. 1(d)), the group velocity can be extracted too.

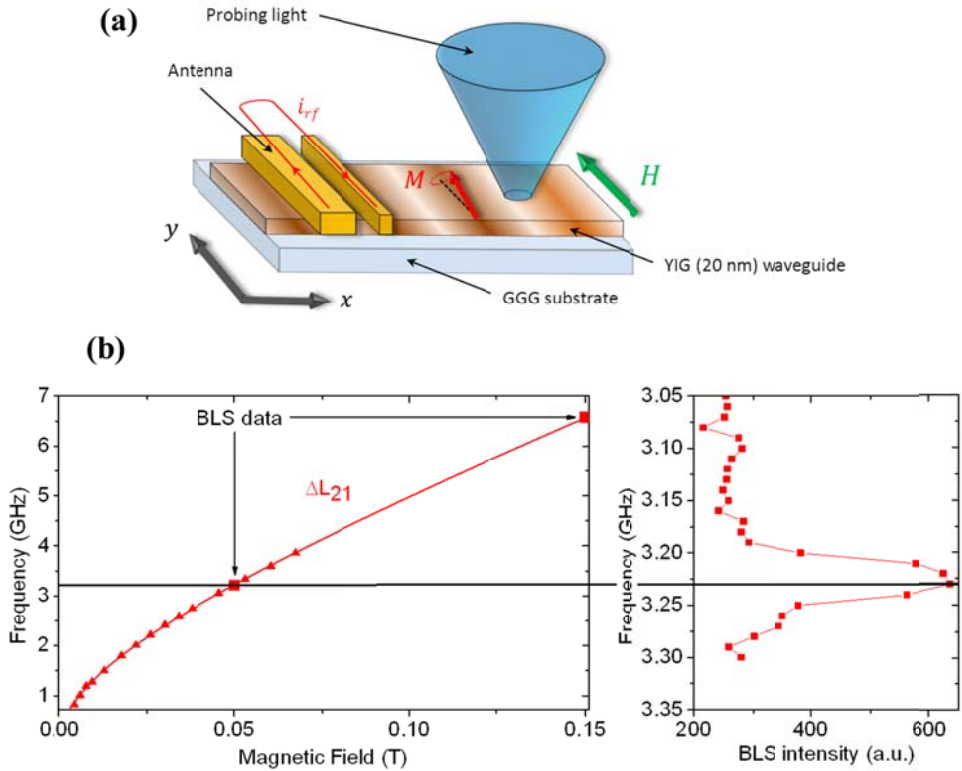

*FIG. 2. (a) Scheme of the BLS setup: the magnetization dynamics of a 2.5 μm wide YIG waveguide is excited by a microwave-field produced by the antenna. The magnetic oscillations are detected via their interaction with a probing light focused directly on YIG. (b) Transmission resonance frequencies extracted for different bias magnetic fields using the PSWS technique (triangles). Line represents a theoretical fit using the Kittel's law equation. Square symbols are obtained using μ-BLS setup (left panel). The right panel shows a typical BLS spectrum recorded on the same YIG waveguide at a 10 μm distance from the excitation antenna at $\mu_0 H$ = 50 mT.*

In summary, the values of the two parameters characterizing the SW propagation at $\mu_0 H = 45$ mT in our 20 nm thick, 2.5 μm wide YIG waveguides are: $L_{att} = 25 \pm 1\ \mu m$ and $V_g = 319 \pm 14$ m/s, from which we can also deduce a magnetization relaxation time $T_2 = L_{att}/V_g = 78 \pm 6$ ns. These experimental values can be compared to the ones obtained from the theoretical modeling of the dispersion relation for a waveguide in the Damon-Eshbach configuration [14,15]. For unpinned spin surfaces and without any quantization along the thickness of the film, the dispersion relation for the $n^{th}$ SW width mode can be expressed as:

$$\omega_n^2 = (\omega_H + \omega_M \Lambda^2 k_n^2 + \omega_M(1 - P))(\omega_H + \omega_M \Lambda^2 k_n^2 + \omega_M P \sin^2 \varphi_n) \qquad (1).$$

Here $\omega_H = \mu_0 \gamma H$ and $\omega_M = \mu_0 \gamma M_{eff}$, with $\mu_0$ the vacuum permeability and $\gamma$ the gyromagnetic ratio, $\Lambda^2 = \frac{2A}{\mu_0 M_{eff}^2}$, with $A$ the exchange constant, and $P = 1 - \frac{1-e^{-k_n d}}{k_n d}$ the dipolar matrix element, with $d$ the thickness of the film. $k_n^2 = k_x^2 + k_{y,n}^2$ and $\varphi_n = \arctan\left(\frac{k_x}{k_{y,n}}\right)$, where $k_{y,n} = \frac{n\pi}{L}$ is the quantized transverse wavevector arising from the lateral boundary conditions ($L$ is the width of the waveguide).

The SWs contributing to the signal are those having wavevectors $(k_x, k_{y,n})$ that match the excitation microwave magnetic field spatial distribution, they can be inferred from the Fourier transform of the spatial profile of the microwave field. However, among those SWs, the one that will contribute most efficiently are those that have large attenuation lengths. By convoluting the k-vector response and the k-vector dependence of the attenuation length we found that the maximum contribution corresponds to $k_x \approx 0.8\ \mu m^{-1}$. Using the expression (Eq. 1) with the magnetic characteristics of our YIG film, we predict theoretically a group velocity $V_g = \frac{\partial \omega}{\partial k} = 350$ m/s and a magnetization relaxation time $T_2 = \alpha \omega \frac{\partial \omega}{\partial H} = 94$ ns, in good agreement with the values found experimentally.

Interestingly, the oscillations observed in Fig. 1(c) are well fitted using a single period, however this is not the case for all excitation frequencies. For example, at 1.4 GHz, we have distinctively observed two oscillation periods. This behavior is indicative of a more complex SW spectrum propagating into the assembly of waveguides. In order to get more insight on this complex behavior, we have performed a μ-BLS study on exactly the same devices used for PSWS. The probing blue laser light produced by a single continuous wave frequency is focused into a diffraction-limited spot on the surface of the YIG film [16,17] (Fig. 2(a)). Using a six-pass Fabry-Perot interferometer, one can analyze the interaction of the probing light with the magnetic excitations in YIG. The resulting BLS signal is proportional to the intensity of the spin waves at the position of the probing spot and at the selected frequency of excitation. For the experiments, a microwave current with low power to minimize the heating is sent through the antenna in a range of frequency that excites propagating SWs in the waveguides. Note that due to instrumentation limitation, most BLS spectra had to be recorded for higher SWs frequencies than the ones for PSWS experiments.

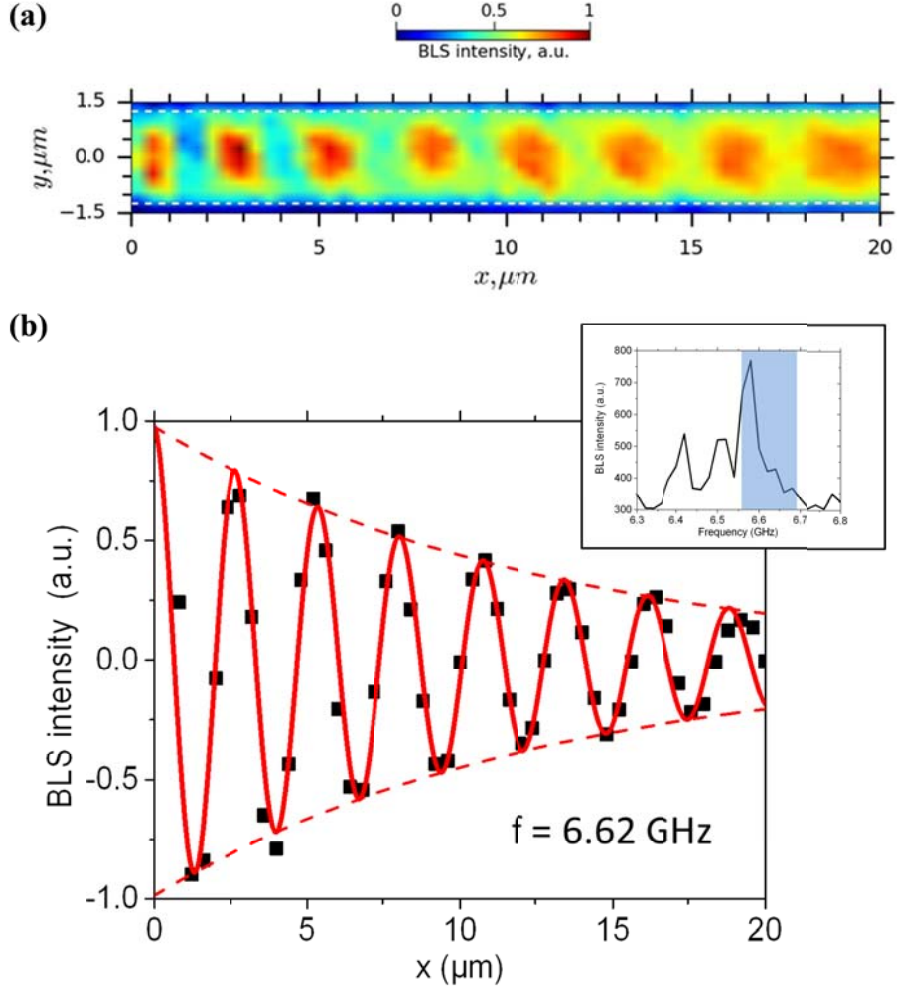

*FIG. 3. (a) Two-dimensional phase map of the propagating spin waves compensated for decay using phase resolved BLS measurements at f = 6.62 GHz over a total area of 3 μm x 20 μm. Dashed lines show the edges of the waveguide. (b) Amplitude profile of the spin wave along the propagation coordinates at f = 6.62 GHz. Red plain line shows the fit of the experimental data. Dashed lines represent the exponential decay of the envelope. The inset shows the excitation frequency spectrum recorded at 500 nm from the antenna. The data were obtained at $\mu_0 H = 0.15$ T.*

On the right panel of Fig. 2(b), we display typical BLS intensity as a function of frequency recorded by shining the laser spot in the middle of a YIG waveguide 10 μm away from the antenna. The BLS spectrum presents one peak at 3.23 GHz which corresponds to the excitation of propagating SWs. Interestingly, the excitation frequency perfectly agrees with the frequencies extracted on $\Delta L_{21}$ data in PSWS measurements (see Fig. 2(b)).

To have access to the phase characteristics of the propagating SWs, we have performed phase resolved μ-BLS experiments [18,19]. In this case, the BLS signal contains information both on the amplitude but also the phase of the propagating SWs. The measurements were performed at high external magnetic field $\mu_0 H = 0.15$ T in order to match the minimum working frequency of the electro-optical modulator used for the phase resolved measurements. As shown in the inset of Fig. 3(b), the antennas allow a wide spectral excitation ranging from 6.4 to 6.7 GHz, we focused our study in the frequency range between 6.56 and 6.68 GHz (blue shaded area) where the phase resolved spectra possess a good coherency. Thus, phase resolved measurements have been performed for different frequencies of the excitation spectrum.

For the spatial phase map of the propagating spin waves presented in Fig. 3(a), microwave pulses with duration of 160 μs are applied to the antenna at a frequency of 6.62 GHz with a power of -20 dBm. The information on the phase is obtained by modulating the BLS signal with a fixed time-delay reference signal. In this case, we observe an interference pattern that reproduces the variation of the SW phase. From these phase sensitive BLS measurements, the wavelength and attenuation length of the spin waves can be extracted directly. Considering a one-dimensional scan along the propagation direction in the middle of a waveguide shown in Fig. 3(b), we can deduce the wavelength corresponding to the distance between two maxima equals λ = 2.7 μm at 6.62 GHz associated to a $k$ = 2.3 μm$^{-1}$. Then the attenuation length can be extracted as follows from the exponential sinusoidal

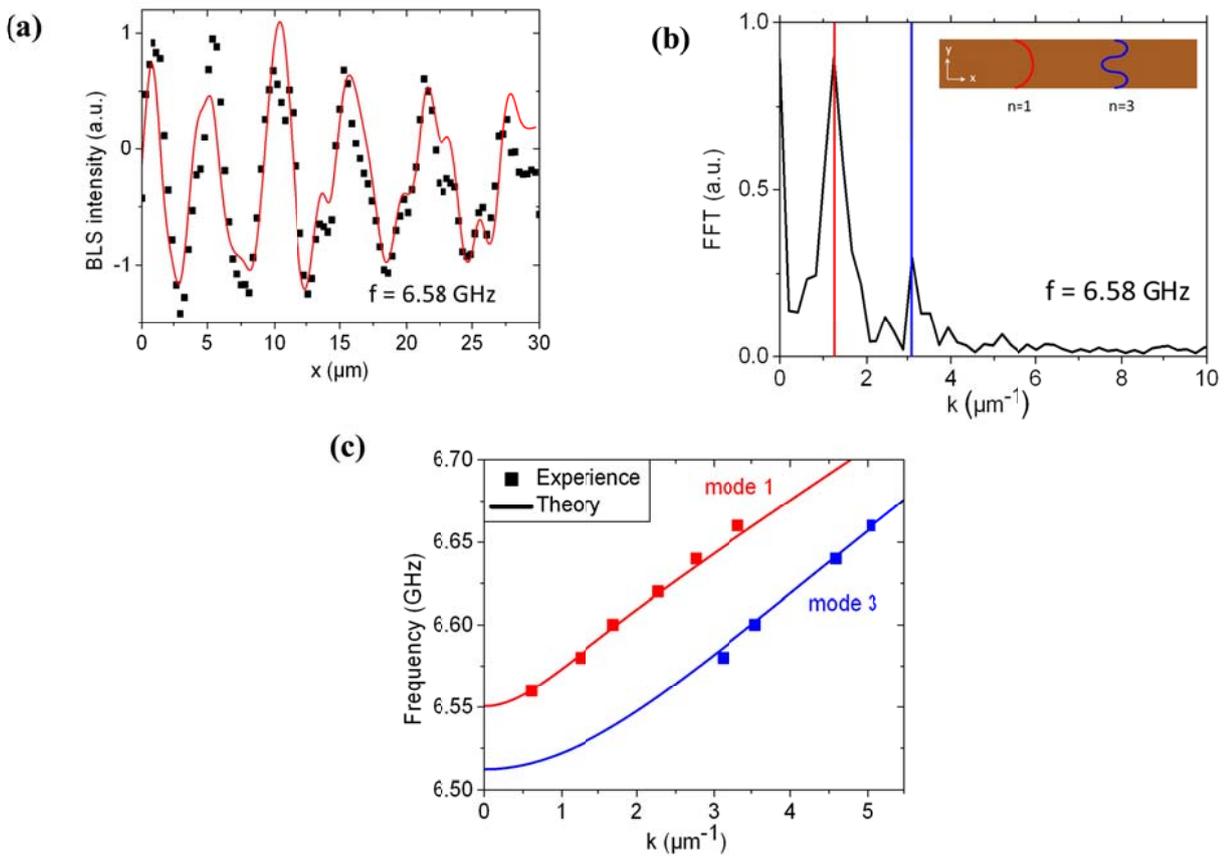

*FIG. 4.(a) Amplitude profile of the spin waves compensated for decay along the propagation coordinates at f = 6.58 GHz. Red plain line shows the fit of the experimental data considering a multi-mode propagation in the waveguide. (b) Fourier transform extracted from the spin-wave signal at f = 6.58 GHz. The two peaks correspond to the first and the third transverse modes in the waveguide. The inset shows the transverse amplitude of these two modes. (c) Square symbols represent experimental dispersion relations using phase resolved measurements for the two peaks extracted from the Fourier analysis signal. Lines show the theoretical dispersion relations for a 2.5 μm wide YIG waveguide for the first two odd modes. The data were obtained at $\mu_0 H$ = 0.15 T.*

waveform [20]:

$$I = A \times \exp(-(x - x_c)/L_{att}) \times \sin(2\pi(x - x_c)/\lambda) \qquad (2)$$

Using Eq. 2, we find that, at this frequency, the attenuation length is $L_{att} = 13 \pm 2$ μm and the signal can be well fitted with a single oscillation period.

For other excitation frequencies, we observe simultaneous excitation of propagating SWs with different wavelengths. Scanning the laser at the center of the waveguide as shown on Fig. 4(a) at $f$ = 6.58 GHz, we observe a more complex SW spectrum. From the Fourier analysis of the line scan displayed in Fig. 4(b), we observe mainly two peaks corresponding to two SW modes dominating the spectrum. Varying the excitation frequency allows us to experimentally reconstruct the dispersion relation i.e. frequency *vs* k-vector, for each of these two main modes. The results are plotted as red and blue symbols in Fig. 4(c). In order to index these two modes, the experimental dispersion relations are compared to the theoretical expected ones deduced from Eq. 1. Taking only the mode index $n$ as a free parameter and keeping all the other ones identical, we identify that the higher-frequency branch corresponds to $n = 1$ and the lower-frequency branch to $n = 3$ (continuous lines in Fig. 4(c)). The absence of a 3[rd] mode at 6.62 GHz can be explained by an extinction in the Fourier transform of the antenna preventing any SWs to propagate at $k \approx 4$ μm$^{-1}$. We emphasize that the absence of mode 2 is expected because of the uniformity of the microwave field along the *y* direction. Similar symmetry based mode selection has already been reported for metallic ferromagnetic stripes [21]. The group velocity is extracted by a linear fit of the experimental dispersion relation. The results are in agreement with theory with a slightly larger group velocity for mode 3 than that of mode 1 ($260 \pm 6$ m/s against $232 \pm 2$ m/s).

In conclusion, we succeed to characterize spin-wave propagation in microfabricated high quality 20 nm thick, 2.5 μm wide YIG waveguides. Spin waves propagation parameters in a wide range of excitation frequency for DE mode configuration have been extracted. As expected from its low damping, YIG allows the propagation of spin waves over large attenuation length (25 μm at $\mu_0 H = 45$ mT and 13 μm for $\mu_0 H = 0.15$ T) in excellent agreement with the theoretical expectations. Direct mapping of spin-waves by μ-BLS allows us to confirm the multi-mode propagation in the waveguide, glimpsed by propagating spin-wave spectroscopy. The electrical detection of SWs in our devices is robust over large propagation distances, up to 70 μm, with a good coherency and amplitude of the signal. The observed good compliance of the SWs with analytical models in such microfabricated YIG waveguides despite the lithographically induced imperfections opens the path to reliable design of complexes magnonics circuits such as logic circuit or microwave application using magnonics crystal.

We acknowledge E. Jacquet, R. Lebourgeois, R. Bernard and A.H. Molpeceres for their contribution to sample growth, and O. d'Allivy Kelly for fruitful discussion. This research was partially supported by the Deutsche Forschungsgemeinschaft and the program Megagrant № 14.Z50.31.0025 of the Russian Ministry of Education and Science. MC acknowledges DGA for financial support. OG thanks IdeX Unistra for doctoral funding.

*abdelmadjid.anane@thalesgroup.com